\documentclass[twocolumn]{article}

\usepackage[dvips]{graphicx}
\usepackage{amsmath}
\usepackage{amssymb}
\usepackage{mathtext}
\usepackage{cite}
\usepackage{multicol}
\usepackage{abstract}

\begin{document}

\title{\bf Activated hopping transport in nematic conducting aerogel at low temperatures}

\author{V.~I.~Tsebro$^{1,2}$\thanks{e-mail: v.tsebro@mail.ru},
E.~G.~Nikolaev$^2$, L.~B.~Lugansky$^2$,
M.~S.~Kutuzov$^3$, \\
R.~A.~Khmel'nitskii$^1$, A.~A.~Tonkikh$^4$, A.~I.~Khar'kovskii\,$^5$
\\
{\small \em $^1$Lebedev Physical Institute, Russian Academy of Sciences, Moscow, 119991 Russia}\\
{\small \em $^2$P.L. Kapitsa Institute for Physical Problems, Russian Academy of Sciences, Moscow, 119334 Russia}\\
{\small \em $^3$Metallurg Engineering Ltd., Tallinn, 11415 Estonia}\\
{\small \em $^4$Prokhorov General Physics Institute, Russian Academy of Sciences, Moscow, 119991 Russia}\\
{\small \em $^5$Institute for High Pressure Physics, Russian Academy of Sciences, Moscow, 142190 Russia}
}

\date{}

\twocolumn[
    \maketitle
    \begin{onecolabstract}
The transport properties of nematic aerogels, which consist of highly oriented Al$_2$O$_3\cdot$SiO$_2$ nanofibers coated with a graphene shell with a large number of defects, are studied. The temperature dependences
of the electrical resistivity in the range of 9--40~K strictly follow the formula derived to describe the variable range hopping (VRH) conductivity, in which exponent $\alpha$ changes from 0.4 to 0.9 when the number of layers in the graphene shell decreases from 4--6 to 1--2. The dependence of $\alpha$ on the shell thickness can be explained by a simultaneous change in the dimensionality of hopping transport and the character of the energy dependence of the density of localized states near the Fermi level. The fact that $\alpha$ approaches unity at the minimum graphene shell thickness indicates a gradual transition from VRH transport to nearest neighbor hopping (NNH) transport. The magnetoresistance measured at T = 4.2~K is negative, increases significantly with decreasing graphene shell thickness, and is approximated by a formula for the case of weak localization with
a good accuracy. The phase coherence lengths are in a reasonable relation with the graphene grain sizes. The conducting aerogels under study complement the well-known set of materials that exhibit hopping electron
transport at low temperatures, which is characteristic of media with strong carrier localization, and also a negative magnetoresistance, which usually manifests itself under weak localization conditions.
    \end{onecolabstract}
]
\saythanks

%\begin{multicols}{2}

\section{Introduction}

Among the entire variety of aerogels, nematic ordered systems are distinguished; they consist of almost parallel to each other nanoscale fibers 5--50~nm in diameter and several centimeters in length  \cite{aghayan2013,asadchikov2015}. The strong anisotropy of such aerogels makes it possible to investigate a number of fundamental physical
phenomena in the media filling an aerogel, as, e.g., was done in works  \cite{dmitriev2015,autti2016} devoted to the quantum properties of superfluid $^3$He. On the other hand, the properties of a nematic ordered aerogel can also be of interest if, e.g., it can conduct an electric current. The authors of the works mentioned above studied
$\gamma$-Al$_2$O$_3$-based aerogels. This material is called nafen
and is manufactured, in particular, by ANF Technology \cite{nafen}. Later, this company developed a similar technology for manufacturing aerogels based on aluminum silicate Al$_2$O$_3\cdot$SiO$_2$ (so-called mullite). The
recently developed CVD technology of carbon deposition
on the surface of nafen/mullite nanofibers makes it possible to fabricate conducting aerogel samples \cite{hussainova2015,ivanov2016,solod2019}. As a result of such graphenization, a shell of several graphene layers with a large number of defects forms on the surface of aerogel nanofibers.

In this work, we study the electrical conductivity of CVD-graphenized mullite samples. By analogy with a number of various disordered carbon-containing systems(in which hopping electrical transport takes place
at low temperatures) investigated earlier \cite{mehbod87,putten92,fung94,reynolds94,mandal97}, we could expect that hopping conduction would also occur in this case and would have peculiar properties due to the specific structure of these aerogels.

The temperature dependence of the electrical conductivity at low temperatures for a hopping mechanism, where variable range hopping (VRH) prevails, is described by the well-known expression \cite{shklovskii79,shklovskii84}
\begin{equation}\label{eq:mott}
    \sigma(T)=\sigma_0\exp \left[ -\left(\frac{T_0}{T}\right)^\alpha\, \right],
\end{equation}
\begin{equation}\label{eq:t0}
    T_0=\frac{\beta}{g(\mu)\xi^d}\ ,
\end{equation}
where $\alpha = 1/(1+d)$, $d$ is the dimensionality of the system,
$g(\mu)$ is the density of localized states at the Fermi level, $\xi$ is the isotropic carrier localization length, and $\beta$ is the numerical coefficient ($\beta$ = 21.2 and 13.8 for $d$ = 3 and 2, respectively). The vast majority of experimental works devoted to the hopping conductivity in various disordered media consider the cases at $d$ = 3 and 2, where the dimensionality of the system according to Eq.~(\ref{eq:mott}) is determined by $\alpha$ in the temperature dependence of conductivity and/or, for a given dimensionality, $\xi$ is determined using Eq.~(\ref{eq:t0}) and $g(\mu)$ found by a certain method in other measurements, or vice versa.

However, when analyzing the experimental data obtained for heavily doped semiconductors, even the authors of earlier works \cite{hill76,zabrod77e} noted that $\alpha$ can have different values from 0.25 to 0.7, although most results in a three-dimensional case were mainly grouped near 0.25. Slightly earlier, the authors of \cite{pollak72,hamilton72} showed that, if the density of localized states at the Fermi level is not constant and changes with energy $\varepsilon$ (measured from the Fermi level) according to the law
\begin{equation}\label{eq:g(e)}
    g(\varepsilon)=g_0|\varepsilon|^n ,
\end{equation}
we have
\begin{equation}\label{eq:alpha(n)}
    \alpha = \frac{n+1}{n+d+1}\, .
\end{equation}
in the case of arbitrary dimensionality $d$.

Therefore, in the case of a strong change in $g(\varepsilon)$ near the Fermi level, $\alpha$ can be significantly higher than 1/4 for $d$ = 3 (or significantly higher than 1/3 for $d$ = 2).

In addition to heavily doped semiconductors, the disordered conducting media exhibiting activated hopping transport at $\alpha$ that differs from the ``classical'' values $\alpha$ = 1/4 and 1/3 (or 1/2) are mainly carbon-containing systems of various structures \cite{mehbod87,putten92,fung94,reynolds94,mandal97}, conducting
polymers \cite{m94,yoon95,aleshin2004}, and polycrystalline
graphene \cite{park2013}.\footnote{The case $\alpha$ = 1/2 is known to be not only a one-dimensional version of Eq.~(\ref{eq:alpha(n)}), but above all to be the result of appearance of a Coulomb gap near the Fermi level due to electron-electron interaction regardless of the dimensionality (Shklovskii-Efros law \cite{shklovskii79}).}
For example, $\alpha$ in carbon black (CB)-polymer composite systems ranges from 0.5 to 0.79 \cite{mehbod87,putten92,mandal97}, which was explained in terms of the model of electronic state superlocalization in a fractal structure \cite{levy87,deutscher87}. According to the results of studying the hopping conduction in polycrystalline graphene \cite{park2013}, specific values $\alpha$ = 0.41 and 0.72 were observed along with the value ($\alpha$ = 0.33) characteristic of 2D systems. The authors \cite{park2013} suggest that, although graphene is a 2D system, conduction in this case occurs via jumping between conducting crystallite boundaries, and its behavior was interpreted using a quasi-one-dimensional model \cite{fogler2004}.

Among the works on conducting polymers, the low temperature transport in which is described by the VRH mechanism \cite{kaiser2001}, we mention work \cite{m94}, where the hopping conduction of a polyaniline (PANI) network
in a polymethylmethacrylate (PMMA) matrix was studied. $\alpha$ was found to decrease from 1 to 0.25 with increasing PANI content above the percolation
threshold. The high values of $\alpha$ in this system were assumed to be associated with the superlocalization of electron wavefunctions due to a fractal character of the PANI network not far from the percolation threshold. At the immediate vicinity of the percolation threshold (where $\alpha \approx 1$), the nearest neighbor hopping (NNH) mechanism takes place. Below the percolation threshold, the behavior of $\sigma (T)$ corresponds to Eq.~(\ref{eq:mott}) with $\alpha$ = 1/2, which is characteristic of granular metals \cite{pollak92}.

A high value $\alpha$ = 0.65--0.70 at $T<50$~K was also observed for polydiacetylene single crystals in \cite{aleshin2004}. This result is explained by quasi-one-dimensional hopping conduction and the influence of the Coulomb interaction. Above $T$ = 50~K, the NNH transport
($\alpha = 1$) dominates.

In highly anisotropic conducting polymer samples, such as PEDOT:PSS thin films, the authors of \cite{nardes2007a} found that $\alpha$ = 0.25 for the lateral direction ($\sigma_\|$) and $\alpha$ = 0.81 for the perpendicular (vertical) direction ($\sigma_\bot$) at a ratio $\sigma_\|/\sigma_\bot$ = 10--10$^3$. This behavior was interpreted as the manifestation of the following two different jumping mechanisms in the lateral and vertical
directions: VRH for $\sigma_\|$ and NNH for $\sigma_\bot$ (despite the fact that $\alpha$ is noticeably lower than unity in the latter case). However, using a numerical simulation, Ihnatsenka \cite{ihnatsenka16} showed that, as the wavefunctions of localized states become anisotropic, $\sigma$ in the direction where the localization length becomes smaller follows
Eq.~(\ref{eq:mott}) in terms of the VRH mechanism. Here, $\alpha$ can
be in the range from 1/4 to 1, which explains the experimental data obtained for PEDOT:PSS \cite{nardes2007a}.

Since the nematic conducting aerogels studied in this work are also highly anisotropic materials, it is interesting to study their transport properties and to reveal a relationship between these properties and the structure of these materials. Electron-microscopic analysis showed that the well-ordered nanofiber structure of nematic aerogels can be represented as a strongly compressed (in the transverse direction) wavy network of conducting nanofibers, the contacts between which exist at the distances much longer
than their diameter. As a result of measuring bulk samples, we found that the anisotropy of the electrical resistivity is in the range $\rho_\bot/\rho_\|$ = 25--40. The temperature dependence of the conductivity of these aerogels in the temperature range from 9 to 40 K rather strictly
follows Eq.~(\ref{eq:mott}) for hopping conductivity in both the
longitudinal (along nanofibers) and transverse directions. Regardless of the direction, $\alpha$ turned out to take different values, from 0.4 to 0.9, depending on the carbon content (graphene shell thickness on the surface
of nanofibers). In this sense, this situation is fundamentally
different from the above-mentioned case of anisotropic conducting polymers, where the presence of two different hopping conduction modes with
strongly different values of $\alpha$ for different directions was established \cite{nardes2007a,ihnatsenka16}. It is clear that, since aerogel nanofibers coated with a conducting shell come into contact with each other at the distances much exceeding their diameter, $\alpha$ is determined solely by transport along the nanofibers and, hence, is independent of
direction.

\begin{figure}[h]
\begin{center}
\includegraphics[width=7.7cm]{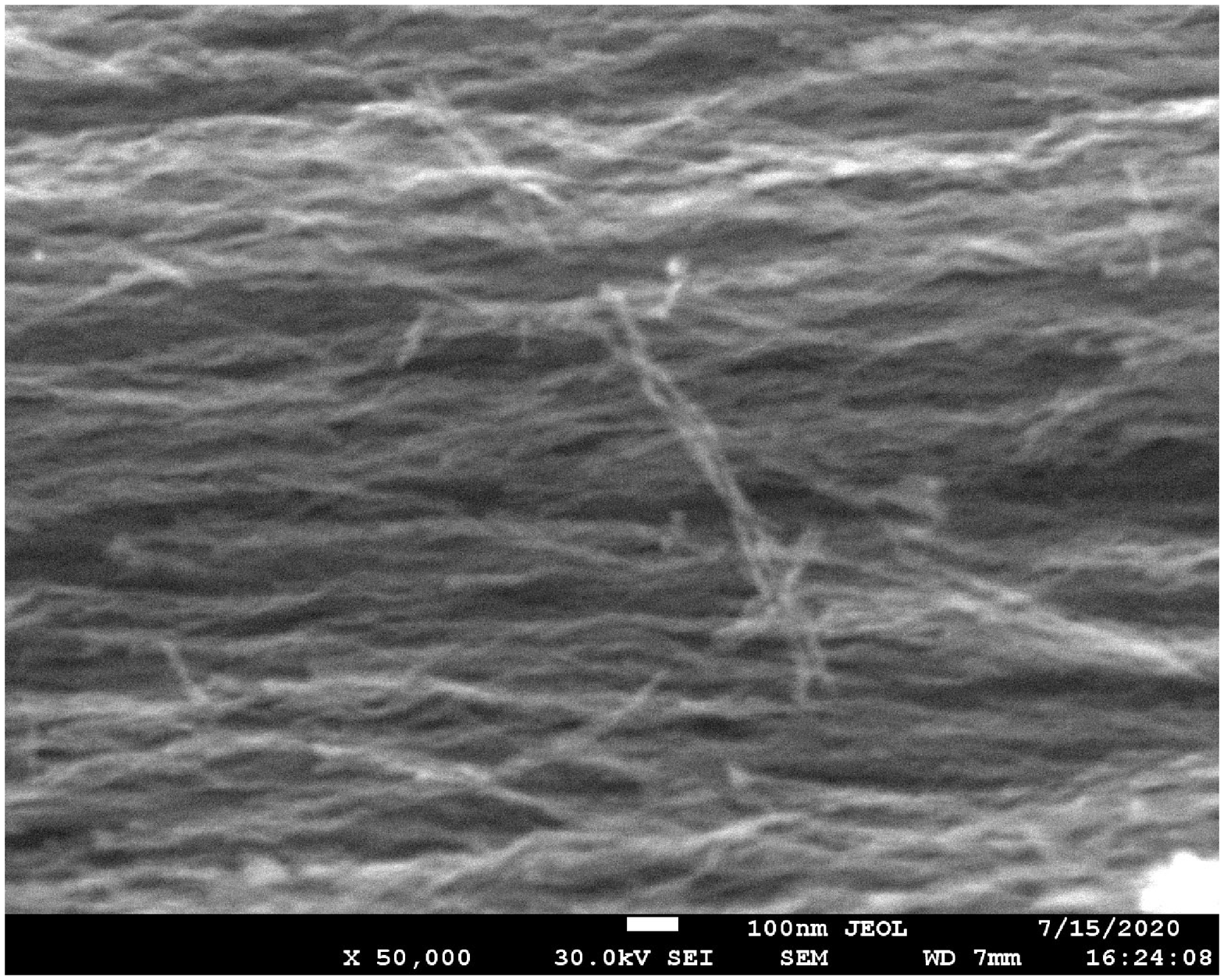}
\includegraphics[width=7.7cm]{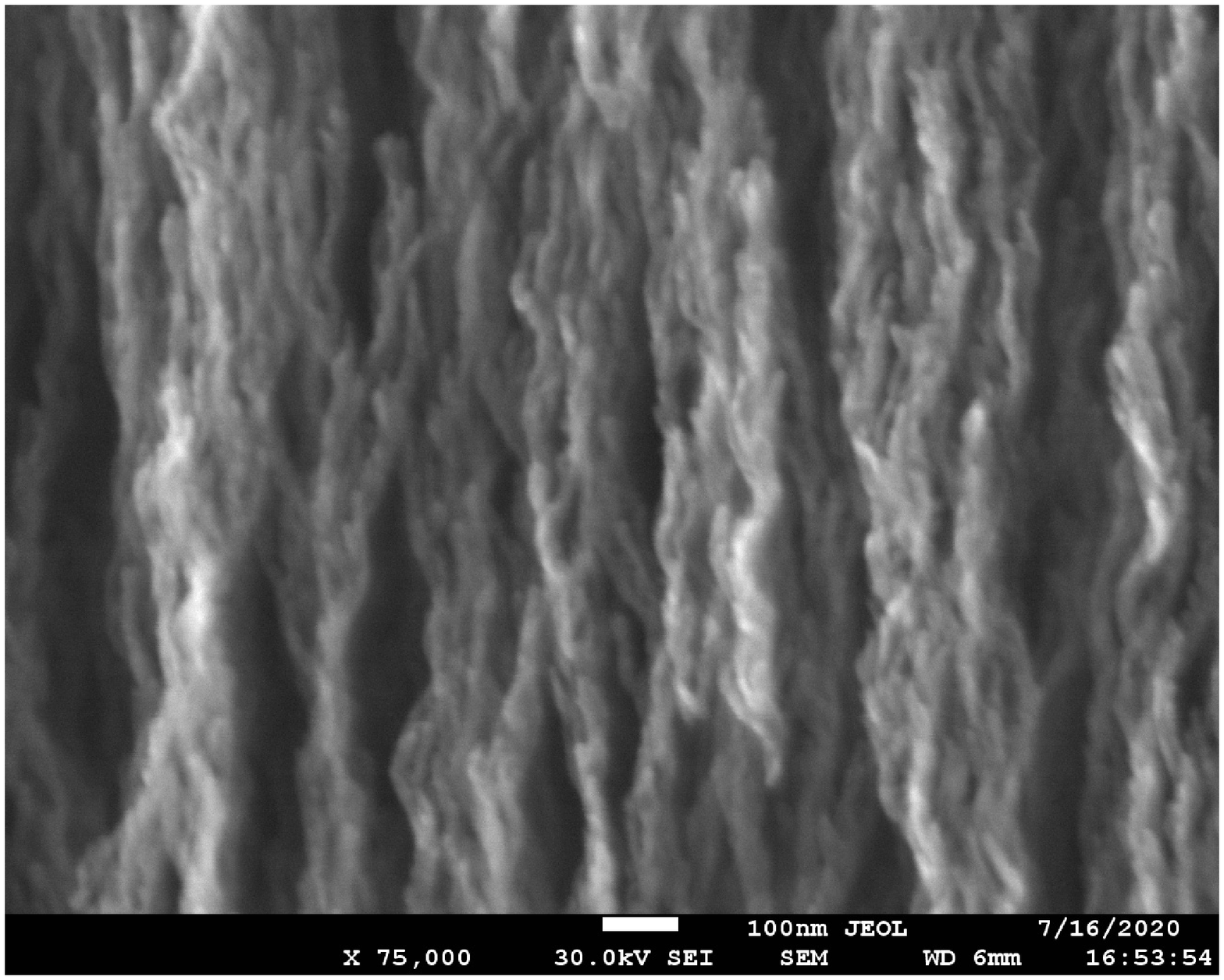}
  \caption{SEM images of the nanofiber structure in aerogel samples (top) AG-14 and (bottom) AG-44.}
  \label{fig:SEM_AG-14-44}
\end{center}
\end{figure}

\section{Aerogel samples and experimental details}

In this work, we studied graphenized samples of mullite, namely, aerogel based on aluminum silicate (Al$_2$O$_3\cdot$SiO$_2$). As noted above, the samples were CVD graphenized similarly to \cite{hussainova2015,ivanov2016,solod2019}. Depending on the CVD time, samples with various carbon contents were prepared, and their values were estimated by XPS during electron microscopy of the samples; the carbon content averaged over many points was taken as the final value. As a result, samples with a carbon content of 14 at.\% (AG-14), 20 at.\% (AG-20), 31 at.\% (AG-31), and
44 at.\% (AG-44) were examined.

\begin{figure}
\begin{center}
\includegraphics[width=7.7cm]{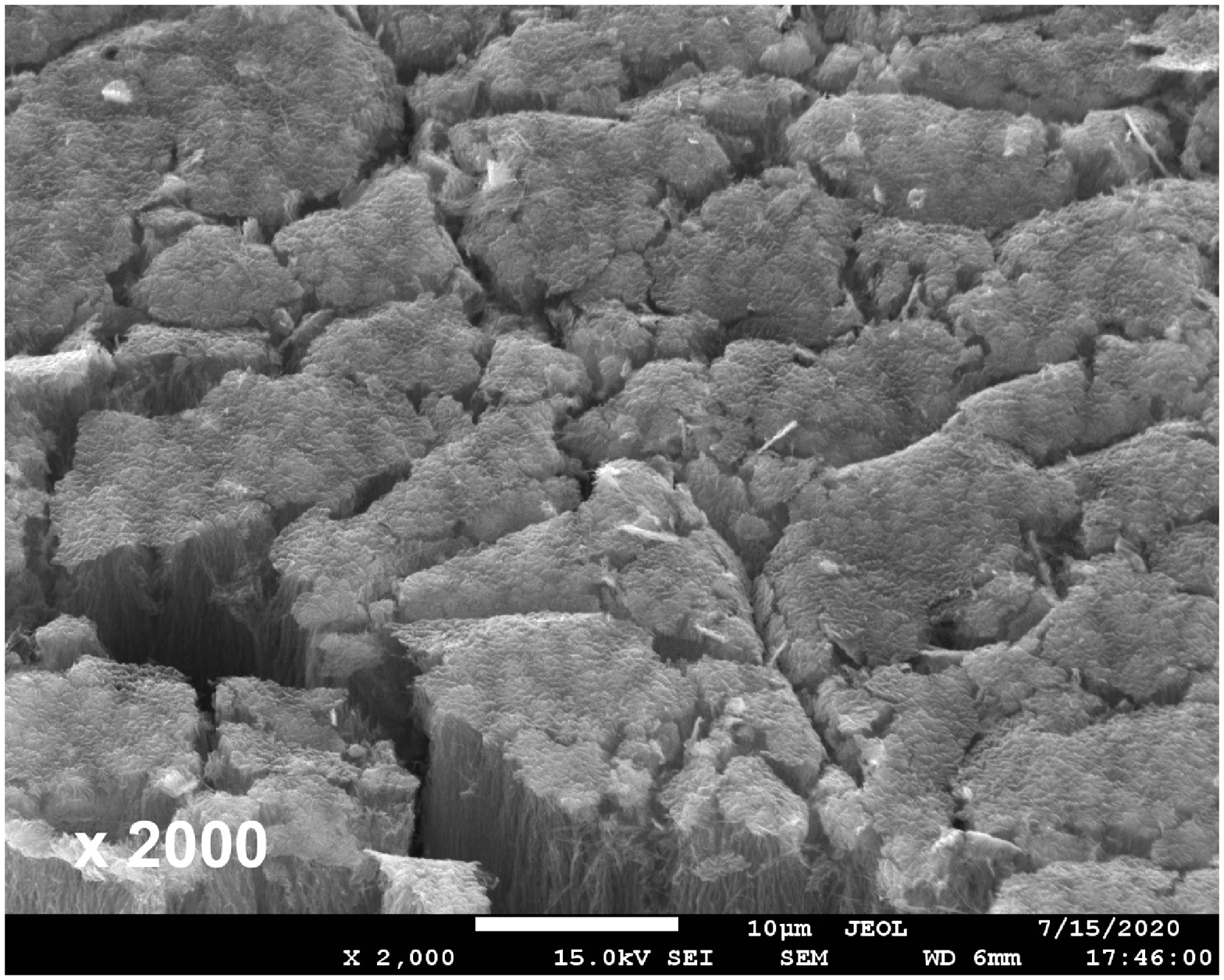}
\includegraphics[width=7.7cm]{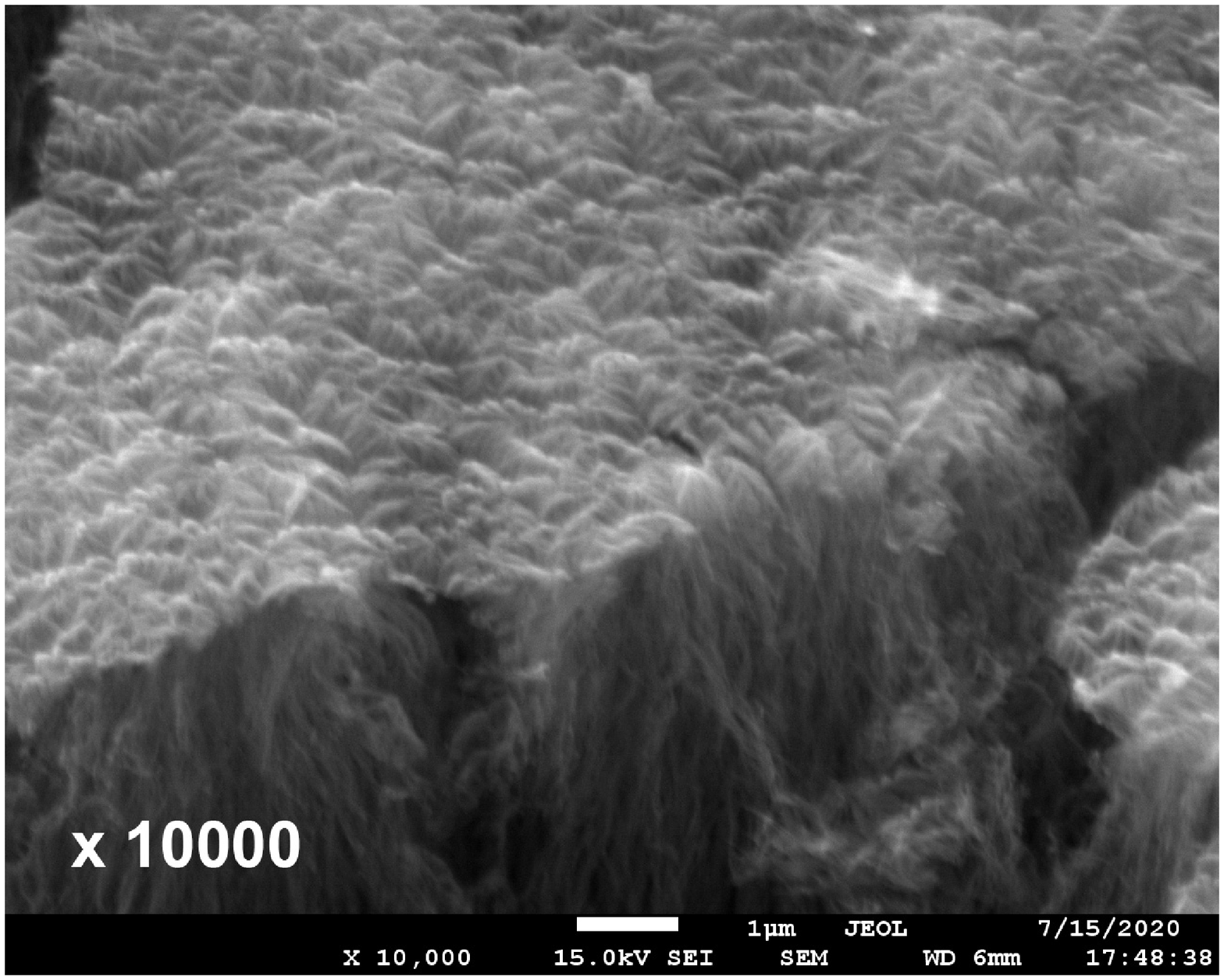}
  \caption{SEM images of the end face of a fragment of the AG-44 aerogel at various magnifications showing a fractal character of the packing of nanofibers.}
  \label{fig:SEM_AG-44}
\end{center}
\end{figure}

The good conductivity of the graphenized samples allowed us to take their images by scanning electron microscopy without using additional conducting
coatings. As can be seen in these images (Fig.~\ref{fig:SEM_AG-14-44}),
wavy and slightly twisted nanofibers 10--15~nm in diameter at the lower scale boundary are located at approximately the same distances between each
other and are almost parallel to each other. A fractal character of the packaging of aerogel nanofibers, which is clearly visible in the images of the end face of a separate nanofiber fragment, is noteworthy. Figure~\ref{fig:SEM_AG-44} shows electron-microscopic images of the end
of such a fragment about 150~$\mu$m in diameter at various magnifications. The cone aggregates clearly visible on the fracture surface at a magnification of 10000 are formed by nanofibers closing in one fractal
group. As can be seen in Figs.~\ref{fig:SEM_AG-14-44} and \ref{fig:SEM_AG-44}, large voids exist between groups of nanofibers assembled in a bundle, and such a picture is reproduced when the scale
is changed by two orders of magnitude.

Using the carbon contents determined by XPS in the samples and the average nanofiber diameter, we estimated the graphene shell thickness at 1--2
graphene layers for the samples with a minimum carbon content (AG-14) and 4--6 layers for the samples with the maximum content (AG-44).

\begin{figure}[h]
\begin{center}
\includegraphics[width=8.0cm]{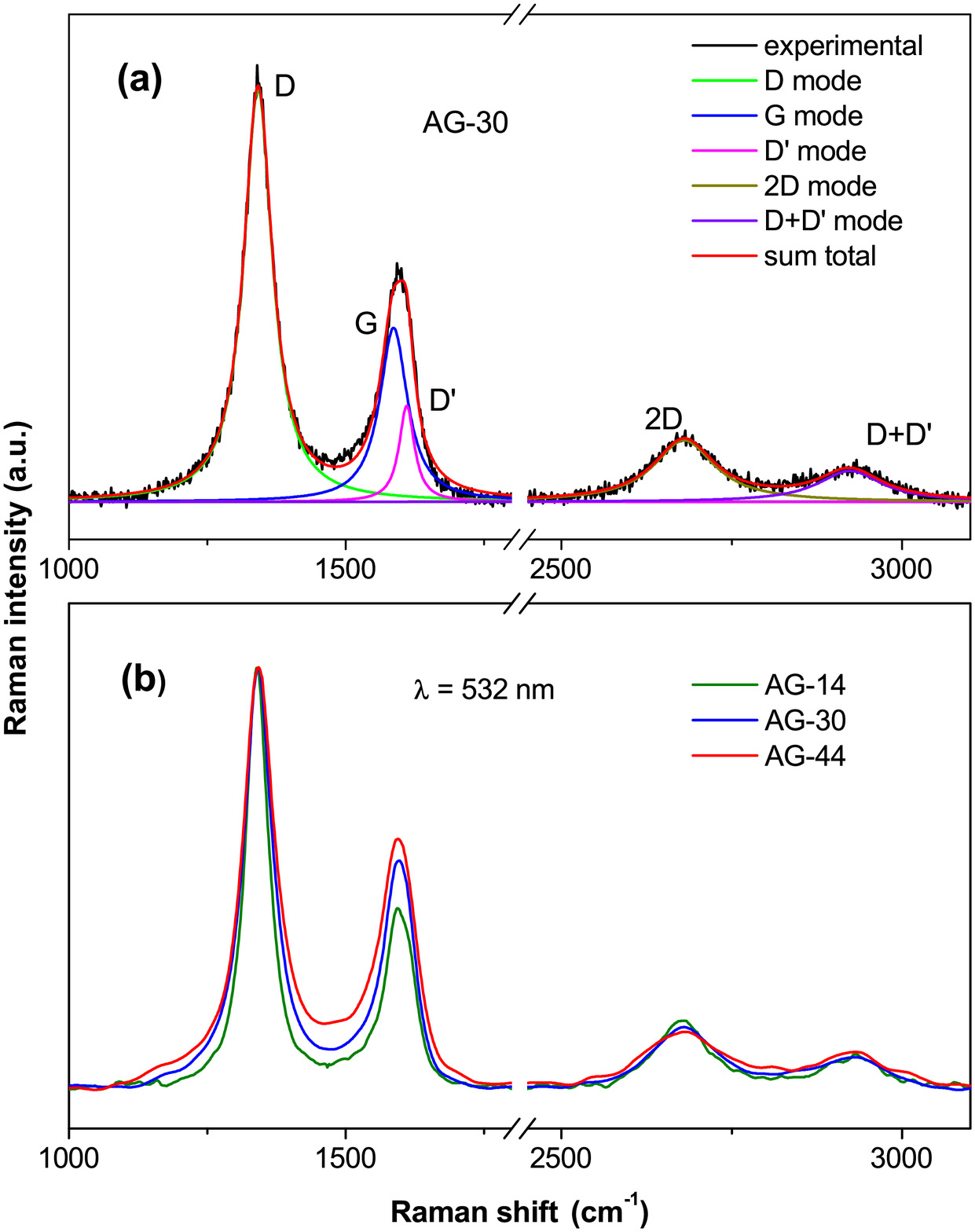}
  \caption{Raman spectra: (a) ((black line) measured spectrum of the AG-30 sample, (colored lines) contributions of individual modes approximated by Lorentzians, (red line) final approximation of the spectrum) and (b) normalized Raman spectra of three aerogel samples with different carbon contents at $\lambda$ = 532 nm.}
  \label{fig:raman}
\end{center}
\end{figure}

Raman spectroscopy (RS) is known to be a powerful tool to characterize a defect state in various graphene-like systems \cite{saito2011,ferreira2010,lucchese2010,dresselhaus2010,eckmann2012,cancado2006,cancado2011}. Figure~\ref{fig:raman} shows the Raman spectra of the aerogel samples under study. These spectra are seen to contain the phonon modes characteristic of the $sp^2$ modifications of carbon (graphite, graphene, carbon nanotubes, etc.) \cite{saito2011}. Spectral features such as the position of modes, their relative intensities, and the peak widths are usually used to determine a specific modification of a graphite-like system, including nanocarbon. In our case, the main characteristic peaks of graphene, namely, the $D$ mode at a frequency of 1340~cm$^{-1}$, the $G$ mode at a frequency of 1588~cm$^{-1}$, and the $2D$ mode at about 2680~cm$^{-1}$, are observed. A high relative intensity of the $D$ mode usually indicates a large number of defects, both single and extended ones \cite{ferreira2010,ivanov2016,lucchese2010,dresselhaus2010}. A low relative intensity of the $2D$ mode is also a marker of significant structure imperfection (or nanodispersity) \cite{eckmann2012}.

\begin{table*}[tb]

  \caption{Results of analyzing the Raman spectra shown in Fig.~\ref{fig:raman} for aerogel samples with different carbon contents: the parameters of the $D$, $G$, $D'$, and $2D$ peaks approximated by Lorentzians; the ratio of the $D$-band to $G$-band intensities ($I_D/I_G$), and average graphene grain size $L_a$ estimated by Eq.~\ref{eq:L_a}}.
  \label{tab:raman}
  \begin{center}
  \begin{tabular}{|r||r|r|r||r|r|r||r|r|r|}
    \hline
    &\multicolumn{3}{c||}{AG-14}&\multicolumn{3}{c||}{AG-30}&\multicolumn{3}{c|}{AG-44}\\
    \hline
    mode&center&width&height&center&width&height&center&width&height\\
    \hline
    D&1339.5&48.9&28.3&1341.5&62.8&27.5&1343.1&77.6&19.1\\
    G&1589.7&54.7&10.9&1586.3&61.0&11.6&1588.0&81.0&10.5\\
    D$'$&1614.5&24.5&4.4&1610.1&33.4&6.4&1608.8&30.6&2.5\\
    2D&2677.6&84.6&4.7&2680.0&108.6&4.1&2681.3&131.5&2.5\\
    D+D$'$&2929.0&99.5&2.1&2923.3&114.4&2.1&2923.6&112.3&1.5\\
    \hline\hline
    $I_{\rm D}/I_{\rm G}$&\multicolumn{3}{c||}{2.6}&\multicolumn{3}{c||}{2.4}&\multicolumn{3}{c|}{1.8}\\
    \hline
    $L_a$(nm)&\multicolumn{3}{c||}{7.4}&\multicolumn{3}{c||}{8.1}&\multicolumn{3}{c|}{10.7}\\
    \hline
  \end{tabular}
  \end{center}
    \end{table*}

To characterize the imperfection of the carbon material of the aerogels and to estimate the average graphene grain size on the nanofiber surface $L_a$, we used the ratio of the $D$-band to $G$-band intensities ($I_D/I_G$) \cite{cancado2006,cancado2011,saito2011}. $L_a$ was estimated by the formula \cite{cancado2006}
\begin{equation}\label{eq:L_a}
    L_a ({\rm nm}) = 2.4\cdot 10^{-10}\lambda^4(I_{\rm D}/I_{\rm G})^{-1},
\end{equation}
where $\lambda$ is the RS excitation wavelength (in our case,
532~nm). As follows from Table~\ref{tab:raman}, the ratios $I_D/I_G$ in
samples AG-14 and AG-30 have similar values, namely, 2.6 and 2.4, respectively, and $I_D/I_G$ for sample AG-44 is 1.8. According to Eq.~(\ref{eq:L_a}), $L_a$ is 7.4~nm for AG-14 and increases to 10.7~nm for AG-44. As follows from Table~\ref{tab:raman}, the $D$, $G$, and $2D$ modes broaden noticeably from sample AG-14 to AG-44, which can be interpreted as a consequence of an increase in the layering of the graphene-like system. However, the number of graphene layers cannot be numerically estimated
because of high structure imperfection.

The electrical resistivities of the aerogel samples were measured using the following two methods: \\
\indent 1) the standard four-probe method on detached material fragments with a large aspect ratio, \\
\indent 2) the modified Schnabel method \cite{lugansky2015} on large bulk centimeter-sized samples having the shape of a rectangular parallelepiped.

\begin{figure}
\begin{center}
\includegraphics[width=8.0cm]{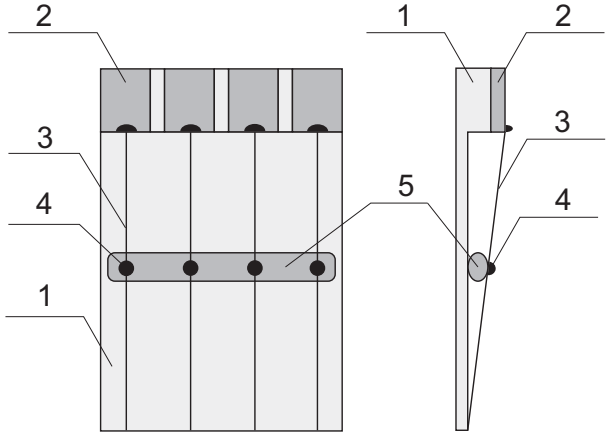}
  \caption{Design of the holder of nanofiber fragment samples: (1) holder base (glass-cloth-base laminate), (2) contact pads, (3) contact wires, (4) self-hardening conducting silver paste, and (5) aerogel sample. }
  \label{fig:holder_str}
\end{center}
\end{figure}

In the first version, the measurements were carried out on detached fragments representing aggregates with a large number of nanofibers. The cross section of such fragments was about 0.05~mm$^2$ at a length of 6--
8~mm. For each sample, an individual special holder was made from thin (0.06--0.08~mm in diameter) tinned copper wires (3) strained at a low angle to holder base plane (1); they served as the leads to current and potential contacts to a sample (Fig.~\ref{fig:holder_str}). After placing a sample (5) in the space between the contact wires and the holder base plane, a small drop of conducting self-hardening silver paste (4) was applied from the back of the contact wires to form a stable contact. It is clear that, in this geometry of the experiment, the measured electrical resistivity is mainly determined by the longitudinal (along the direction of nanofibers) component of the resistivity $\rho_\|$.

Using the second version of $\rho$ measurements by the modified Schnabel method \cite{lugansky2015} on bulk parallelepiped aerogel samples we were able to determine the absolute values of the resistivity components in
both the longitudinal ($\rho_\|$) and transverse ($\rho_\bot$) directions.
The well-known Schnabel method \cite{schnabel64,schnabel67} proposes a solution for a sample in the shape of a planeparallel conducting plate of a given thickness and width and infinite length. In this case, two point
contacts on one side of the sample and two on the opposite side are located at the center exactly opposite each other. The modification of this method described by us earlier in methodical work \cite{lugansky2015} consists in
finding a method to solve the problem of finding the electrical resistivity $\rho$ of an anisotropic sample of finite dimensions. The main features of using this method, as applied to our case, are described in the Appendix.

It should be noted that, before graphenization, the mullite samples are sufficiently strong objects, which can be subjected to mechanical treatment to acquire the required sizes and shape if necessary. However, after graphenization, they become extremely delicate and break up into separate
fragments upon a weak mechanical action. This is a consequence of the fact that the bond between graphenized mullite nanofibers becomes very weak
after they are covered with a graphene shell. Nevertheless, we were able to solve the problem of achieving the necessary sizes and shape for such a high-porosity conducting material and the problem of forming reliable
time-stable electrical point contacts in the right places.

\begin{figure}[h]
\begin{center}
\includegraphics[width=8.0cm]{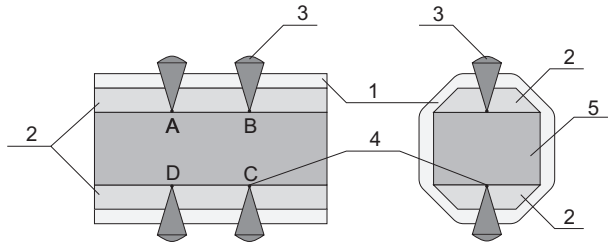}
  \vspace{1cm}
  \caption{Design of the contact assembly for measuring the two components of the electrical resistivity according to the modified Schnabel method: (1) heat-shrinkable shell, (2) cover plates with contacts, (3) conical contacts, (4) self-hardening conducting silver paste, and (5) bulk aerogel sample. $A$, $B$, $C$, and $D$ are the point contact designations used in the Appendix.}
  \label{fig:holder_shnab}
\end{center}
\end{figure}

Bulk samples AG-14 1$10\times3\times2.7$~mm$^3$ in size, AG-30 ($14\times7\times6$~mm$^3$) and AG-44 ($9.8\times3\times2.7$~mm$^3$)
were prepared for measurements. For each sample, an individual contact assembly was made with the sizes determined by the bulk aerogel sizes. The design of the assembly is shown in Fig.~\ref{fig:holder_shnab}. In this design, aerogel sample (5) was placed between two cover plates (2) with conical holes for placing copper conical contacts (3). Before placing a contact in the cover plate, a small amount of self-hardening silver paste (4) was placed on its tip. As a result, the transverse size of the point contact to the sample did not exceed 0.1~mm. The contact assembly was placed in heat-shrinkable shell (1) in order to hold the entire construction and to ensure its integrity and the necessary strength.

Assemblies with samples were mounted on a bulk copper thermoblock, which was placed in a chamber with heat-exchange gaseous helium at a pressure of 5--10~Torr in a cryostat for intermediate temperatures. The temperature dependences of conventional resistances $R_1(T)$ and $R_2(T)$ (see Appendix) were measured upon very slow heating of the thermoblock and switching the measuring current to exclude thermal and contact emf. The heating rate was changed from 0.05~K/min near $T$ = 5~K to 0.2 K/min near $T$ = 40~K. Despite the low heating rates, the samples were supercooled due to the desorption of the heat-exchange gas during measurements on large bulk samples in the temperature range from 20 to 26~K because of the poor thermal conductivity of the high-porosity aerogel. Therefore, the data measured in this range were excluded from consideration (see Fig.~\ref{fig:W(T)_rho3_AG-14} below). No supercooling of samples in the heat-exchange gas desorption area was
observed during measurements on AG fragments with small cross-sectional area (Figs.~\ref{fig:W(T)_four_stripes}, \ref{fig:lnR(T)_AG-20}).

\section{Results and discussion}

The temperature dependences of the conductivity of the aerogels were measured in the temperature range 5--50~K, since just in this range they follow Eq.~(\ref{eq:mott}) with various values of $\alpha$ and $T_0$. For data processing to determine the values of $\alpha$ and $T_0$, we used the approach proposed in \cite{zabrod77e,zabrod84e} for the systems where
VRH transport is observed in a relatively small temperature range. In this approach, the temperature dependence of the logarithmic derivative of the conductance $W(T)$ is plotted in logarithmic coordinates,
\begin{equation}\label{eq:W(T)}
    W(T)=T[d\ln\sigma(T)]/dT=\Delta\ln\sigma(T)/\Delta\ln T \ .
\end{equation}
The linear section in the $W(T)$ dependence determines the temperature range in which Eq.~(1) is strictly fulfilled, and the required parameters $\alpha$ and $T_0$ are found from the equations
\begin{equation}\label{eq:logW(T)}
\begin{array}{l}
    \log W(T)=A-\alpha \log T \\
    A=\alpha \log T_0 + \log \alpha \ .
\end{array}
\end{equation}

\begin{figure}[h]
\begin{center}
\includegraphics[width=8.0cm]{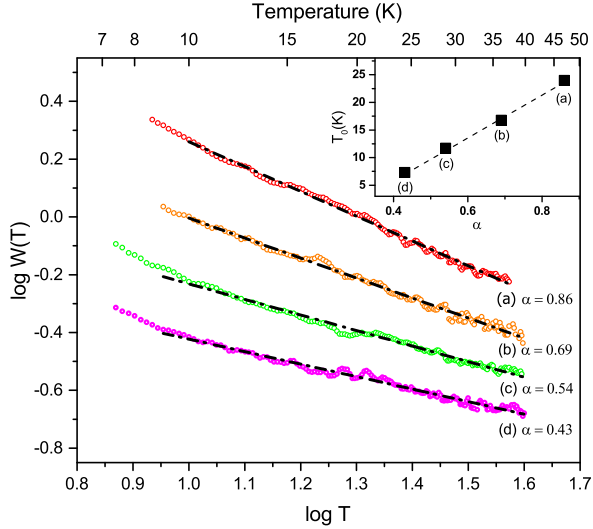}
  \caption{Temperature dependences of the logarithmic derivative of the conductance for four nanofiber fragment samples with various carbon contents: (a) AG-14, (b) AG-20, (c) AG-31, and (d) AG-44. (inset) $T_0$ vs. $\alpha$.}
  \label{fig:W(T)_four_stripes}
\end{center}
\end{figure}

\begin{figure}
\begin{center}
\includegraphics[width=8.0cm]{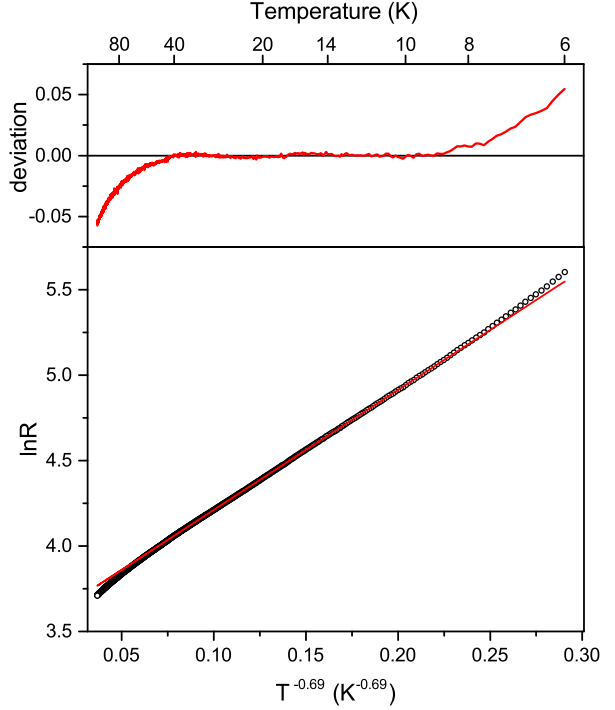}
  \caption{Nanofiber fragment sample AG-20: (a) logarithm of the resistance vs. $T^{-0.69}$ and (b) deviation of experimental points from a linear dependence.}
  \label{fig:lnR(T)_AG-20}
\end{center}
\end{figure}

Figure~\ref{fig:W(T)_four_stripes} shows the temperature dependences of the
logarithmic derivative of conductance
\begin{equation*}
    W(T)=\Delta\ln G(T)/\Delta\ln T ,
\end{equation*}
where $G=R^{-1}$, for four AG fragments with different carbon contents. As follows from these data, $\alpha$, which is determined by the slopes of the linear sections in Fig.~\ref{fig:W(T)_four_stripes} according to Eq.~(\ref{eq:logW(T)}), increases monotonically when the graphene shell thickness of the aerogel nanofibers decreases (from sample AG-44 to sample
AG-14). Table~\ref{tab:aT0_str} gives parameters $\alpha$ and $T_0$ for all four samples. The relationship between these parameters has a well-pronounced linear character (see the inset to Fig.~\ref{fig:W(T)_four_stripes}). Note that the temperature dependence of
the conductivity strictly follows Eq.(\ref{eq:mott}) for all samples in the temperature range 9--40 K, which is illustrated as an example by the data shown in Fig.~\ref{fig:lnR(T)_AG-20} for sample AG-20, i.e., the dependence of $\ln R$ on $T^{-0.69}$ and the deviation of experimental points from a linear dependence.

\begin{table}
  \centering
  \caption{Parameters $\alpha$ and $T_0$ for the following four nanofiber fragment samples: AG-14, AG-20, AG-31, and AG-44}

  \label{tab:aT0_str}
  \begin{tabular}{|c||c|c|}
    \hline
          & $\alpha$ & $ T_0$, K \\ \hline
    AG-14 & 0.86 & 24.0 \\
    AG-20 & 0.69 & 16.8 \\
    AG-31 & 0.54 & 11.7 \\
    AG-44 & 0.43 & 7.3 \\
    \hline
  \end{tabular}
\end{table}

\begin{figure}
\begin{center}
\includegraphics[width=8.0cm]{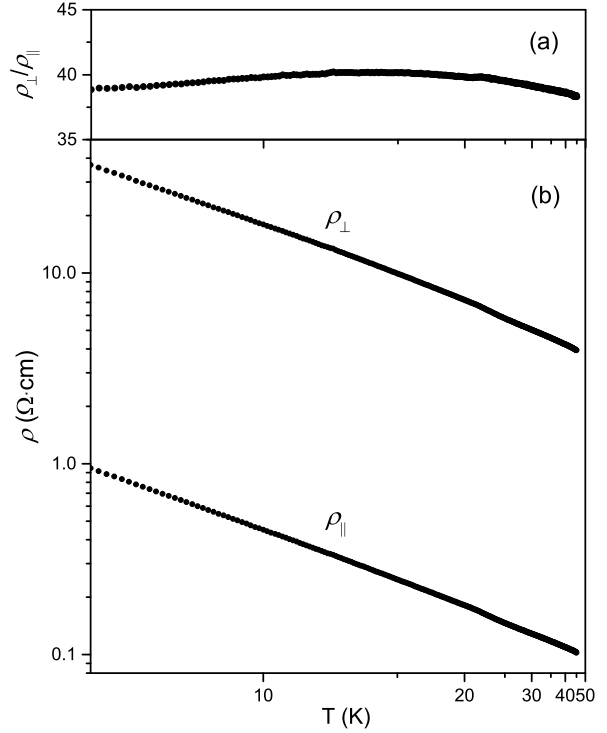}
  \caption{Temperature dependences of (a) the anisotropy of the electrical resistivity and (b) its two components $\rho_\|$ and $\rho_\bot$ measured by the Schnabel method for bulk sample AG-14.}
\label{fig:anisotropy_AG-14}
\end{center}
\end{figure}

As noted above, the temperature dependences of electrical resistivity were measured on large bulk parallelepiped aerogel samples AG-14, AG-31, and
AG-44 using the modified Schnabel method. Figure~\ref{fig:anisotropy_AG-14}
shows the temperature dependences of the longitudinal ($\rho_\|$) and transverse ($\rho_\bot$) resistivity components in logarithmic coordinates and the anisotropy of the conducting medium ($\rho_\bot/\rho_\|$) for bulk sample AG-14. Both components are seen to be strongly different and to change with temperature in a similar way, and the anisotropy depends weakly on temperature.

\begin{figure}
\begin{center}
\includegraphics[width=8.2cm]{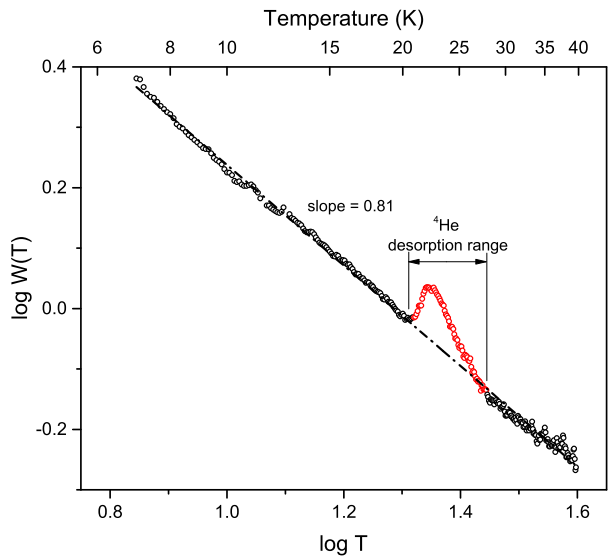}
  \caption{Temperature dependences of the logarithmic derivative of the transverse conductivity component for bulk sample AG-14.}
\label{fig:W(T)_rho3_AG-14}
\end{center}
\end{figure}

As an example of processing the data obtained on bulk samples, Fig.~\ref{fig:W(T)_rho3_AG-14} depicts the temperature dependences
of the logarithmic derivative of the transverse conductivity component,
\begin{equation*}
    W(T)=\Delta\ln \sigma_\bot(T)/\Delta\ln T
\end{equation*}
for sample AG-14. The data are seen to fall on a straight line, the slope of which determines $\alpha_\bot$ = 0.81, except for the temperature range where the sample is supercooled because of the desorption of the heat-exchange gas (from 20 to 26 K).

Table~\ref{tab:aT0_bulk} gives the values of parameters $\alpha$ and $T_0$ for both conductivity components of the bulk aerogel samples of all three compositions under study, and Table~\ref{tab:ro_bulk} presents the absolute values of the electrical resistivity and anisotropy at two temperatures at the beginning and the end of the temperature range where the temperature dependence of conductivity follows Eq.~\ref{eq:mott}.

\begin{table}
  \centering
  \caption{Parameters $\alpha_\|$, $\alpha_\bot$, $ T_0^\| $, and $ T_0^\bot $ for bulk samples AG-14, AG-31, and AG-44}
  \label{tab:aT0_bulk}
  \begin{tabular}{|c||c|c|c|c|}
    \hline
          & $\alpha_\| $ & $\alpha_\bot $ & $ T_0^\|$, K & $ T_0^\bot $, K \\ \hline
    AG-14 & 0.90 & 0.81 & 22.6 & 27.6 \\
    AG-31 & 0.58 & 0.58 & 25.7 & 27.1 \\
    AG-44 & 0.49 & 0.49 & 5.8 & 6.9 \\
    \hline
  \end{tabular}
\end{table}

\begin{table}
  \centering
  \caption{$\rho_\|$, $\rho_\bot$, and anisotropy $\rho_\bot/\rho_\|$ of bulk samples AG-14, AG-31, and AG-44 at temperatures of 4.5 and 45~K}
  \label{tab:ro_bulk}
  \begin{tabular}{|c||c|c|c|c|}
    \hline
          & $T$, K & $\rho_\|$, $\Omega\cdot$cm & $\rho_\bot$, $\Omega\cdot$cm & $\rho_\bot/\rho_\|$ \\ \hline
    AG-14 & 4.5 & 3.28 & 130.7 & 39.8 \\
                        & 45 & 0.103 & 3.94 & 38.3 \\
    \hline
    AG-31 & 4.5 & 4.99 & 137.5 & 27.5 \\
                        & 45 & 0.58 & 15.3 & 26.3 \\
    \hline
    AG-44 & 4.5 & 0.81 & 39.0 & 48.2 \\
                        & 45 & 0.35 & 15.9 & 45.4 \\
    \hline
  \end{tabular}
\end{table}

\begin{figure}
\begin{center}
\includegraphics[width=8.0cm]{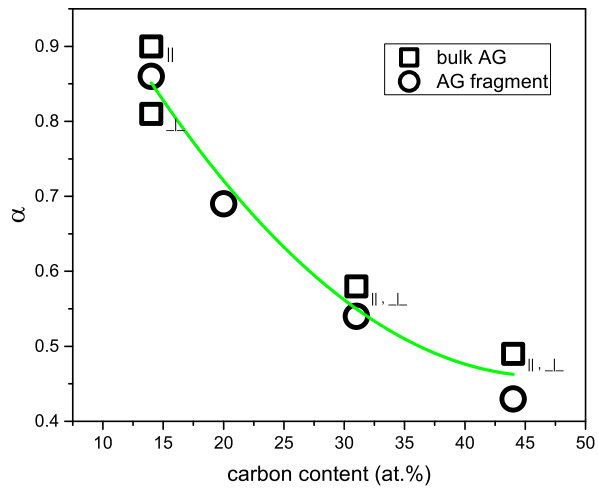}
  \caption{$\alpha$ vs. the carbon content.}
\label{fig:alpha_C}
\end{center}
\end{figure}

Figure~\ref{fig:alpha_C} summarizes the values of $\alpha$ obtained on four AG fragments and on the bulk samples for both conductivity components. All data are seen to group around a general dependence of $\alpha$ on the carbon content. The values of $\alpha_\|$ and $\alpha_\bot$ in the bulk samples
exactly coincide with each other except for sample AG-14, where a certain difference between these values takes place. This difference includes the value of $\alpha$ for the sample in the form of a long-sized fragment (it
is clear that $\alpha_\|$ is mainly measured in this case). The equality of the values of $\alpha_\|$ and $\alpha_\bot$ for the bulk samples indicates that $\alpha$ is determined by transport on the graphene shell, along the surface of aerogel nanofibers. The anisotropy of the effective bulk conductivity measured by the modified Schnabel method is determined
by the morphology of the fibrous structure, namely, the intersection of nanofibers with each other at the distances significantly exceeding their diameter.

Thus, the nematic conducting aerogels (graphenized mullite) represent an
extremely rare case of a system, in which $\alpha$ changes monotonically as a function of a single parameter, namely, the carbon content or, in other words, the graphene shell thickness of aerogel nanofibers, at an unchanged morphology.\footnote{In this regard, work \cite{m94} (mentioned above), where $\alpha$ increased from 0.25 to 1 when the volume fraction of a PANI network in a PMMA matrix decreased, is noteworthy. In addition, according
to \cite{park2013}, a certain set of $\alpha$ values was observed in a series of polycrystalline graphene samples.} To explain this dependence, we have to assume that Eq.~(\ref{eq:alpha(n)}) is also valid in our case and, when the graphene shell thickness increases, $d$ and $n$ in Eq.~(\ref{eq:alpha(n)}) change from $d$ = 1 and $n$ = 2 (AG-14) to $d$ = 3 and $n$ = 1 (AG-44). If this is the case, a quasi-one-dimensional VRH mechanism should be considered for the thinnest graphene shells. Theoretically, such a consideration for a single one-dimensional chain and many interacting parallel chains was made, e.g., in \cite{fogler2004} (so-called FTS model), where $d$ = 1, 2, and 3 and $n$ = 0, 1, and 2 were noted. Based on the FTS model, the authors of \cite{park2013} interpreted the experimental data on hopping conduction in semiconductor polycrystalline graphene with a crystallite size from several to several tens of nanometers. The quasi-one-dimensional hopping transport of carriers was assumed to occur due to jumps between conducting crystallite boundaries (edges). The existence of such conducting boundaries (mainly zigzag edges) was
experimentally shown in \cite{ritter2009}.

In our case, a one-dimensional hopping conduction mode can occur if the following circumstance is taken into account. For the sample with the minimum carbon content (coating thickness of 1--2 layers), the average graphene grain size according to RS results is 7--8 nm. In this case, the nanofiber diameter turns out to be comparable with the grain size. If the motion of charge carriers in such a system is considered as jumps between grains (or, as noted above, between conducting grain boundaries), it is obvious that the hopping transport in this case should have a predominantly
one-dimensional character. When the carbon coating thickness increases, the motion of charge carriers along the graphene shell, apparently, can no longer be considered as one-dimensional. At the same time, as follows from the data on negative magnetoresistance given below, the total imperfection of the system increases despite an increase in the average grain size.

The high value of $\alpha$ for the samples with the minimum carbon content can mean that the system is approaching transition from the VRH to NNH. A similar case was observed in \cite{m94} on studying a PANI network in a PMMA matrix, where $\alpha$ increased to unity when the PANI content decreased up to the percolation threshold. As was found in that work, the NNH mechanism in the system under study was realized when the diameter of PANI filaments decreased to the value comparable with the characteristic jump distance.

In contrast to our case and the work mentioned above, the authors of \cite{fung94}, where carbon aerogel samples with conducting chains of carbon granules were studied, found that the transverse size of the conducting
chains remained unchanged but their length increased when the aerogel density decreased. That system was thought to be described by the granular
metal model \cite{pollak92}, and the fact that $\alpha$ = 1/2 for all samples was attributed to the presence of a Coulomb gap. The transition to one-dimensional carrier motion occurs when the characteristic jump distance becomes smaller than the average length of the conducting chains without their intersection (average chain length between nodes), when the temperature increases.\footnote{Another carbon system, namely, so-called carbynes, should be mentioned here; they are supposed to consist of chains of carbon atoms with the $sp$ bonds. Hopping conductivity with $\alpha$ = 1/2, 1/3, and 1/4 is observed in carbyne samples depending on the synthesis temperature, and the case of $\alpha$ = 1/2 is explained by a one-dimensional character of hopping transport rather than by the presence of a Coulomb gap (see \cite{demishev2002} and Refs. therein).} The transition from VRH to NNH occurs when the jump distance becomes smaller than the double granule size.

The linear relationship detected between $\alpha$ and $T_0$ in Eq.~(\ref{eq:mott}) (see the inset to Fig.~\ref{fig:W(T)_four_stripes}) is noteworthy. This relationship means that the dependences of $\alpha$ and $T_0$ on the effective dimensionality have a similar shape. As far as we know, a simultaneous increase in $\alpha$ and $T_0$ on changing the system parameters was not noted in any of the previously published works. The only exception is paper \cite{mandal97}, where a correlation between $\alpha$ and $T_0$ was detected for two groups of samples with CB of different
origins in a CB--polymer composite system when the CB concentration changed. In this case, the values of $\alpha$ and $T_0$ in each group weakly depended on the CB concentration, and it was not specified what exactly is the difference between these two groups, and the causes of the correlation were not discussed.

\begin{figure}[h]
\begin{center}
\includegraphics[width=8.0cm]{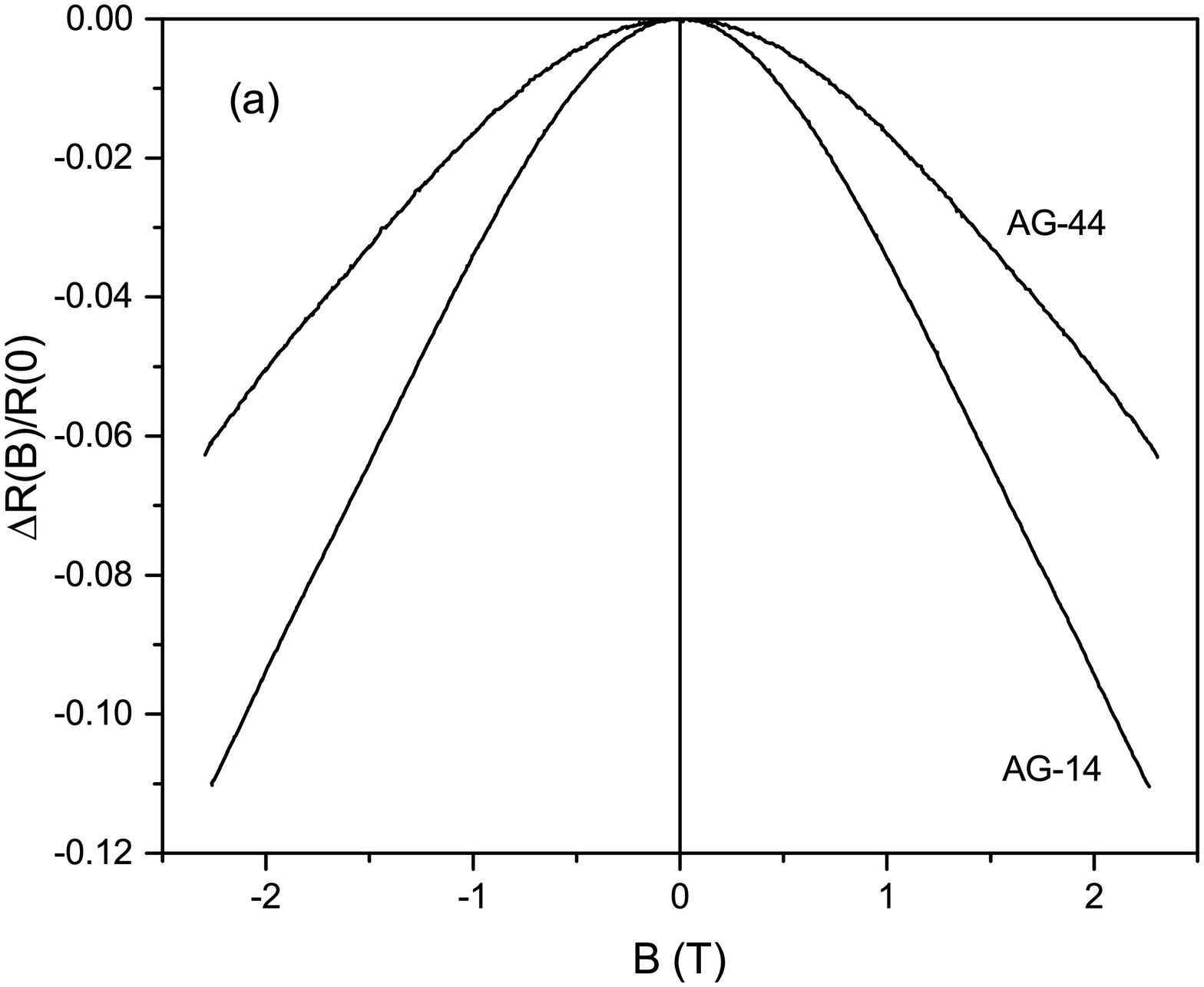}
\includegraphics[width=8.0cm]{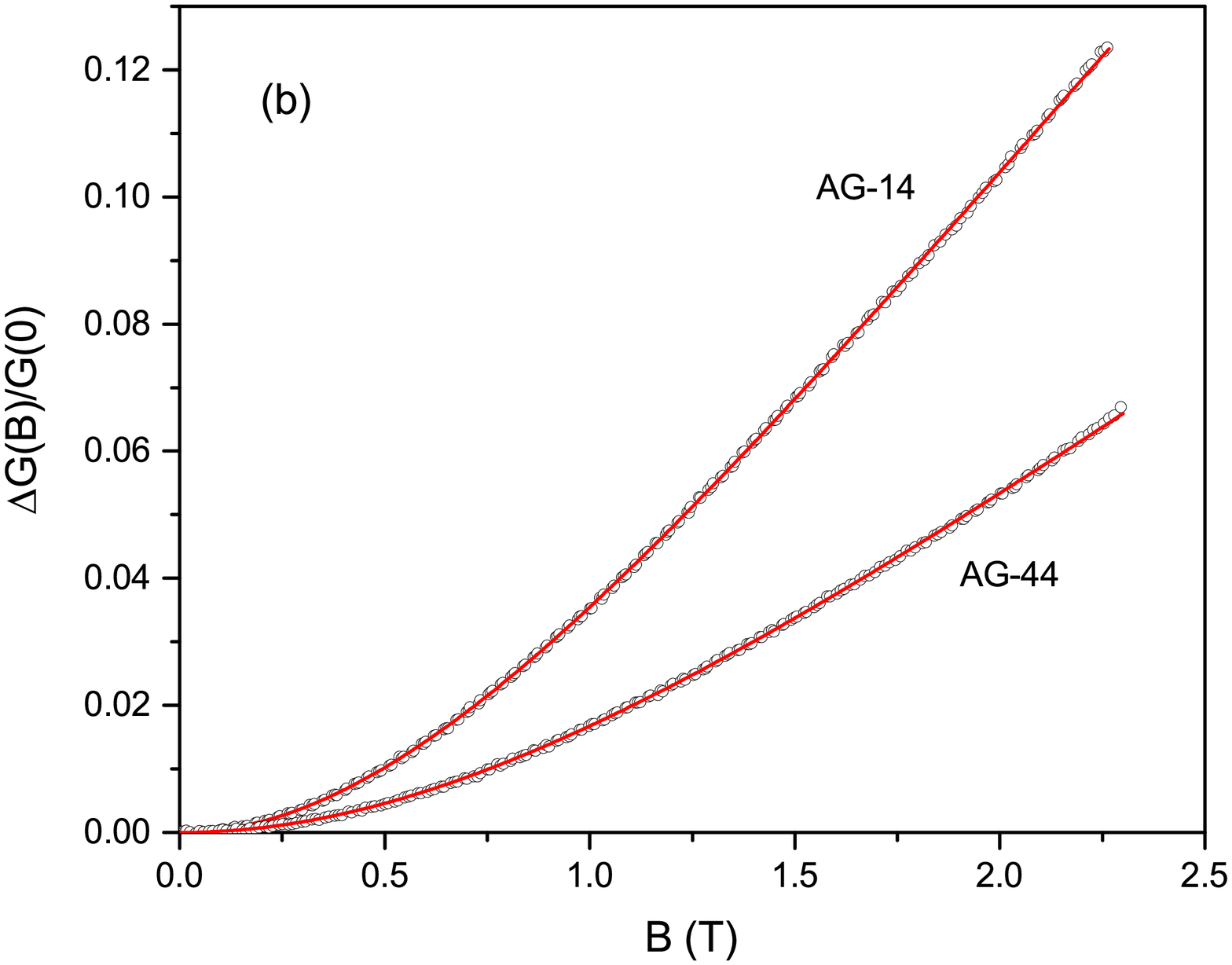}
  \caption{AG-14 and AG-44 samples: (a) reduced dependences of magnetoresistance $\Delta R(B)/R(0)$ and (b) approximations of the experimental dependences of magnetoconductance $\Delta G(B)/G(0)$ by Eq.~(\ref{eq:MC(B)}). $T$ = 4.2 K.}
\label{fig:R(H)_MC(H)}
\end{center}
\end{figure}

The rather low values of $T_0$ should be noted (see Tables~\ref{tab:aT0_str},\ref{tab:aT0_bulk}). Especially for the samples with a high carbon content, they are near the lower limit of the temperature range (9--40 K) where the temperature dependence of conductivity follows Eq.~(\ref{eq:mott}). This fact can indicate a large carrier localization length $\xi$. In a large number of cases for systems with the VRH conduction mechanism, $\xi$ can be estimated using magnetoresistance measurements, since, according to VRH theory \cite{shklovskii79,shklovskii84}, the magnetoresistance is positive (because of the field-induced compression of the wavefunction of localized carriers) and (in not a very strong magnetic fields)
\begin{equation*}
    \ln [\rho(B)/\rho(0)]\propto (\xi/L_B)^4(T_0/T)^{3/(d+1)}
\end{equation*}
($L_B$ is the magnetic length). Positive magnetoresistance in systems with a strong carrier localization and the VRH hopping transport mechanism was observed in a variety of disordered media, which allowed $\xi$ to be
determined. It was also of interest in our case to measure the magnetoresistance behavior of the aerogel samples in not very strong magnetic fields.

Such measurements at $T$ = 4.2~K in magnetic fields up to 2.4~T were performed on aerogel samples in the form of long-sized fragments, since the magnetoresistance measured in this geometry is predominantly transverse, when a current flows through a sample perpendicular to a magnetic field vector. Figure ~\ref{fig:R(H)_MC(H)}a shows the measured dependences $\Delta R(B)/R(0)$ for samples AG-14 and AG-44. The magnetoresistance in our case is seen to be negative, to have a significant magnitude, and to decrease considerably with increasing carbon content.

As is known (see, e.g., \cite{lee85}), a negative magnetoresistance is characteristic of systems with weak carrier localization in the diffusive transport regime, where $k_Fl\gg1$ ($k_F$ is the quasi-Fermi momentum, $l$ is the carrier path length) and $\sigma(T)\propto\ln(T)$. This effect is caused by dephasing in the magnetic field of self-intersecting trajectories of the diffusion motion of carriers in the forward and reverse directions. However, there are numerous cases where negative magnetoresistance is observed in various systems and in the case of a strong localization with VRH mechanism, when the temperature dependence $\sigma(T)$ at low temperatures follows Eq.~(\ref{eq:mott}). Among these systems, we note the GaAs/AlGaAs \cite{jiang92} and GaAs/InGaAs \cite{minkov2002} heterostructures, Ge films \cite{mitin2007}, fluorinated graphene \cite{hong2011}, In$_2$O$_{3-x}$ films \cite{milliken90,frydman95}, highly disordered carbon nanofibers \cite{wang2006}, and netlike films made of single-layer carbon nanotubes \cite{wang2018}.

Theoretically, the mechanism of negative magnetoresistance in the case of strong carrier localization and the VRH transport mechanism was proposed in
well-known work \cite{nguen85} (and then improved in \cite{sivan88,entin89,ioffe2013}). The interference between all possible various hopping trajectories from one localization center to
another was shown to significantly change the carrier tunneling probability depending on conditions (so called NSS model). A negative magnetoresistance
appears in this model as a result of the destruction of this interference by a magnetic field, similarly to how it occurs in systems with weak localization in the diffusion mode. Different magnetic-field dependences of
magnetoconductance $G$ were obtained: in not very strong magnetic fields, $\Delta G(B) \propto B$ B in \cite{nguen85} while $\Delta G(B) \propto B^2$ in \cite{sivan88,entin89,ioffe2013}.

However, among the experimental works mentioned above (\cite{jiang92,minkov2002,mitin2007,hong2011,milliken90,frydman95,wang2006,wang2018}), we would like to note those (see, e.g., work \cite{wang2006} on carbon nanofibers and \cite{wang2018} on nanotube networks) where the field dependence of magnetoconductance is well described by the following
expression, which is typical of systems with weak localization and diffusion electron transport, over a wide magnetic field range \cite{lee85}:
\begin{equation}\label{eq:MC(B)}
    \Delta G(B)= A(T)\left[\Psi\left(\frac{1}{2}+\frac{B_{\phi}}{B}\right)+
    \ln\left(\frac{B}{B_{\phi}}\right)\right] ,
\end{equation}
where $A(T)$ is a temperature-dependent constant determined by the system parameters, $\Psi$ is the digamma function, and $B_{\phi}=\hbar/4eL_{\phi}^2$ ($L_{\phi}$ is the phase coherence length).

Equation~(\ref{eq:MC(B)}) turned out to describe very well the field dependences of magnetoconductance for aerogel samples with different carbon contents in our case as well (see Fig.~\ref{fig:R(H)_MC(H)}b, where the results of such an approximation are shown). The approximation results give
$B_{\phi}$ = 0.745~T for AG-14 and 1.012~T for AG-44, from whence we have $L_{\phi}$ = 14.9 and 12.8~nm, respectively. When comparing these values with the average graphene size $L_a$ obtained from the optical RS data (see Table~\ref{tab:raman}), we can see that $L_{\phi}$ for the samples with
the minimum graphene shell thickness of aerogel nanofibers (AG-14) is about twice as large as $L_a$, whereas $L_{\phi}$ for the maximum shell thickness (sample AG-44) only slightly exceeds La. Note also that, according to the optical data, $L_a$ increases with the graphene shell thickness, and $L_{\phi}$, on the contrary, decreases slightly. Thus, if the use of Eq.~\ref{eq:MC(B)} to describe the negative magnetoresistance of the aerogel samples is valid, it means that an increase in the effective dimensionality of the system with the number of graphene layers in a nanofiber shell leads to a decrease in the phase coherence length. In other words, although the average graphene grain size increases with the shell thickness, the total imperfection of the system also increases.

The use of Eq.~(\ref{eq:MC(B)}) to describe the negative magnetoresistance
in our case is justified not only by the fact that it results in reasonable values of the electronic parameters of the systems under study (as was shown in \cite{wang2006,wang2018}), but also by special experiments \cite{minkov2002}. Using a single quantum well in the GaAs/In$_x$Ga$_{1-x}$As/GaAs heterostructure with a single two-dimensional carrier-filled subband of dimensional quantization as an example, the authors of \cite{minkov2002} showed that the quantum corrections to conductivity are rather significant not only in the diffusion mode at $k_Fl=$2--25, but also at significantly smaller free path lengths in the range $k_Fl=$0.5--2. Note also work \cite{tarki2004} on transport in highly disordered multilayer carbon nanotubes, where the behavior of magnetoresistance as a function of field follows Eq.~(\ref{eq:MC(B)}) with a good accuracy at $l<10$~nm.

\section{Conclusion}

When studying the transport properties of nematic conducting aerogels, we found that the temperature dependence of the electrical conductivity of these materials in the temperature range 9--40~K strictly follows
Eq.~(\ref{eq:mott}) for VRH transport. The conducting medium in which this transport occurs is the graphene shell of aerogel fibers, and its thickness is the main parameter that affects $\alpha$ in Eq.~(\ref{eq:mott}) and the negative magnetoresistance. When the number of layers in the carbon shell decreases from 4--6 to 1--2, $\alpha$ increases from 0.4 to 0.9. According to Eq.~(\ref{eq:alpha(n)}), the increase in $\alpha$ to 0.75 can be explained by a decrease in the dimensionality of the system and an increase in exponent $n$ in the energy dependence of the density of localized states
near the Fermi level. This behavior is consistent with the fact that, at the minimum carbon content, the nanofiber diameter becomes comparable with the
graphene grain size. As a result, hopping transport acquires a predominantly one-dimensional character. A further increase in $\alpha$ up to 0.9 actually means that the system is approaching a transition to NNH transport.
Transport measurements in a magnetic field showed that the magnetoresistance of carbon-coated aerogel samples is negative and decreases significantly
with increasing carbon content. The observed dependence $R(H)$ was approximated with a good accuracy by Eq.~(\ref{eq:MC(B)}) for the case of weak localization. The estimated phase coherence lengths are in a reasonable relation with the graphene grain sizes in the carbon coating of fibers. Thus, the aerogels studied in this work complement the well-known set of systems that exhibit hopping electron transport at low temperatures, which is characteristic of media with strong carrier localization, and, at the same time, a negative magnetoresistance, which usually manifests itself
during diffusion transport under weak localization conditions.

\vspace{20pt}

\noindent {\bf Funding.} This work was supported by the Russian Science Foundation,
project no.~RNF-20-42-08004.

\section*{\raggedleft Appendix}

In the Schnabel method, two measurement methods are possible. In the first Schnabel geometry, the electric current flows through contacts $A$ and $D$ and potential difference $V_{BC}$ between contacts $B$ and $C$ is measured (see the designations of the contacts in the left part of Fig.~\ref{fig:holder_shnab}). In the second Schnabel geometry, the
current flows through contacts $A$ and $B$ and potential difference $V_{CD}$ is measured between contacts $C$ and $D$. From such measurements, conventional resistances $R_1=V_{BC}/I_{AD}$ and $R_2=V_{CD}/I_{AB}$ are determined. $V_{BC}$ and $V_{CD}$ can be analytically expressed by solving the problem of the electric field potential distribution in the sample volume when the current $I$ passes through the corresponding contacts.

In Schnabel's original works \cite{schnabel64,schnabel67}, this problem
was solved for a sample in the shape of an infinite flat plate, where only two geometric parameters, namely, plate thickness $d$ and the distance between neighboring contacts $s$, were present, and in the shape of an
infinite strip, where another geometric parameter, namely, strip width $b$, was added.

In our works \cite{lugansky2015,lugansky2015b}, a solution to this problem
was found for samples having the shape of a rectangular parallelepiped of finite dimensions. $R_1$ and $R_2$ were shown to be represented as
\begin{equation*}
    R_1=\frac{\rho}{d}\:G(a,\,b,\,d,\,s), \quad R_2=\frac{\rho}{d}\:H(a,\,b,\,d,\,s).
\end{equation*}
where $\rho$ is the electrical resistivity of the conducting medium; $a$, $b$, and $d$ are the sample sizes along the principal axes; and $s$ is the distance between contacts $AB$ and $CD$, respectively. Functions $G$ and $H$ are analytically expressed in the form of double infinite series
\cite{lugansky2015,lugansky2015b}. In the case of an isotropic sample, resistivity $\rho$ can be found from any one of these measurements
(either from $R_1$ or $R_2$).

In an anisotropic case, van der Pauw \cite{pauw61} showed that a simple linear transformation of coordinates can be used to reduce the problem of potential distribution in an anisotropic sample to a similar problem for a
hypothetical isotropic sample with different sizes and electrical resistivity. Here, we briefly present the main final calculations concerning the case of the anisotropic nematic aerogel samples studied in this work.

The coordinate system is assumed to be chosen so that the edges of the bulk aerogel samples are along the principal axes of the resistivity tensor, taken as axes $(x_1, x_2, x_3)$, and segments $AB$ and $CD$ are assumed to be parallel to axis $x_1$ along aerogel nanofibers (Fig.~\ref{fig:holder_shnab}). In this coordinate system, tensor $\rho_{ik}$ is diagonal and has only three components ($(\rho_1, \rho_2, \rho_3)$). The coefficients of the linear coordinate transformation are chosen so that the electrical resistivity of an isotropic sample $\rho^*$ and the sizes $a^*, b^*, d^*, s^*$ are
\begin{multline*}
\rho^*=(\rho_1\rho_2\rho_3)^{1/3},\\
a^*=\left(\rho_1/\rho^*\right)^{1/2}a,\qquad b^*=\left(\rho_2/\rho^*\right)^{1/2}b,\\ d^*=\left(\rho_3/\rho^*\right)^{1/2}d,\qquad
s^*=\left(\rho_1/\rho^*\right)^{1/2} .
\end{multline*}
Here superscript * denotes the quantity related to the isotropic image of a real anisotropic sample. In this case, measured resistances $R_1$ and $R_2$ of the anisotropic sample under study are equal to the corresponding
resistances of its hypothetical isotropic image, $R^*_1$ and $R^*_2$.

As is seen from the analytical formulas for functions $G$ and $H$ (see \cite{lugansky2015,lugansky2015b}), they actually depend on three rather than four ($a, b, d, s$) arguments; it is convenient represent these three arguments as ratios ($(a/s,\,b/s,\,d/s)$) for the sample under study and ($(a^*/s^*,\,b^*/s^*,\,d^*/s^*)$) for its isotropic image. Then resistances $R_1$ and $R_2$ to be measured can be written as
\begin{multline*}
R_1=R^*_1=\frac{\rho^*}{d^*}\:G(a^*/s^*,\ b^*/s^*,\ d^*/s^*)=\\
=\frac{(\rho_1\rho_2)^{1/2}}{d}\:G(a/s,\ \lambda_{21}b/s,\ \lambda_{31}d/s)
\end{multline*}
\begin{multline*}
R_2=R^*_2=\frac{\rho^*}{d^*}\:H(a^*/s^*,\ b^*/s^*,\ d^*/s^*)=\\ \frac{(\rho_1\rho_2)^{1/2}}{d}\:H(a/s,\ \lambda_{21}b/s,\ \lambda_{31}d/s),
\end{multline*}
where the designations $\lambda_{21}=(\rho_2/\rho_1)^{1/2}$ and $\lambda_{31}=(\rho_3/\rho_1)^{1/2}$ are introduced.

Only two quantities, namely, $R_1$ and $R_2$, are independent during measurements. Therefore, it is impossible to determine all three values of the resistivity tensor from these measurements. However, if two of the three principal values of the resistivity tensor are the same (as in our case of bulk nematic aerogel samples), then it has only two independent
principal values, which can be found by measuring $R_1$ and $R_2$. Note that, in our experiments on bulk aerogel samples, the lines ($AB$ and $CD$) along which the probes are located are directed along the highest
conductivity direction, which is taken as axis $x_1$ with electrical resistivity $\rho_1$ (see Fig.~\ref{fig:holder_shnab}), and the electrical
resistivities along other two axes are taken to be the same, i.e., $\rho_2=\rho_3$ and $\lambda_{21}=\lambda_{31}$.

Knowing the analytical expressions for functions $G$ and $H$ \cite{lugansky2015,lugansky2015b}, the sample sizes ($a, b, d$), and the distance between point contacts $s$, we can construct the ratio
\begin{equation*}
\frac{R_1}{R_2}=\frac{G(a/s,\ \lambda\,b/s,\ \lambda\,d/s)}
{H(a/s,\ \lambda\,b/s,\ \lambda\,d/s)}
\end{equation*}
as a function of only one argument $\lambda = \lambda_{21}=\lambda_{31}$. For each measurement of $R_1$ AND $R_2$, this dependence is used to determine anisotropy parameter $\lambda$, and $\rho_1$ is then calculated by the formula
\begin{multline*}
\rho_1=\frac{R_1d}{\lambda\,G(a/s,\ \lambda\,b/s,\ \lambda\,d/s)}\\ \mbox{or}\quad
\rho_1=\frac{R_2d}{\lambda\,H(a/s,\ \lambda\,b/s,\ \lambda\,d/s)}\ ,
\end{multline*}
Finally, we find $\rho_3=\rho_1\lambda^2$.

\end{document}